\documentclass[checkin,showpacs,aps,prb]{revtex4}
\newcommand{\ba}{\begin{eqnarray}}
\newcommand{\ea}{\end{eqnarray}}

\usepackage{graphicx,color}

\newcommand{\be}{\begin{equation}}
\newcommand{\ee}{\end{equation}}

\newcommand{\et}{{\it et al. }}

\def\prl{{ Phys. Rev. Lett. }}

\begin{document}

\title{Manifestation of intra-atomic 5d6s-4f exchange coupling in
  photoexcited gadolinium} \author{G. P. Zhang$^*$, T. Jenkins, and
  M. Bennett} \affiliation{Department of Physics, Indiana State
  University, Terre Haute, Indiana 47809, USA} \author{Y. H. Bai}
\affiliation{Office of Information Technology, Indiana State
  University, Terre Haute, Indiana 47809, USA}

\date{\today}

\begin{abstract}
{ Intra-atomic exchange couplings (IEC) between $5d6s$ and $4f$
  electrons are ubiquitous in rare-earth metals and play a critical
  role in spin dynamics. However, detecting them in real time domain
  has been difficult. Here we show the direct evidence of IEC between
  $5d6s$ and $4f$ electrons in gadolinium.  Upon femtosecond laser
  excitation, $5d6s$ electrons are directly excited; their majority
  bands shift toward the Fermi level while their minority bands do the
  opposite. For the first time, our first-principles minority shift
  now agrees with the experiment quantitatively.  Excited $5d6s$
  electrons lower the exchange potential barrier for $4f$ electrons,
  so the $4f$ states are also shifted in energy, a prediction that can
  be tested experimentally. Although a significant number of $5d6s$
  electrons, some several eV below the Fermi level, are excited out of
  the Fermi sea, there is no change in the $4f$ states, a clear
  manifestation of intra-atomic exchange coupling.  Based on our
  results, we propose that the demagnetization time of a material be
  inversely proportional to the density of states at the Fermi level
  and the excited, not the whole, spin moment. This can be tested
  experimentally.  }
\end{abstract}

 \maketitle

\section{Introduction}

Gadolinium (Gd) is the only rare-earth metal that belongs to a group
of elementary ferromagnets for magnetic storage devices \cite{kryder},
and is one of the well studied ferromagnets both experimentally and
theoretically
\cite{scott,moon,erskine,wu,kurz,sabiryanov,petersen2006}.  Gd is also
the key element for all-optical spin switching
\cite{stanciu,prb2017a,jpcm17a,jpcm17b}.  One key feature of
rare-earth metals is that their $4f$ states, deep below the Fermi
energy, are highly localized but contribute a major part of the spin
moment. In Gd, 7$\mu_B$ out of 7.55 $\mu_B$ are from half-filled $4f$
electrons \cite{drulis}. Thus Gd could be considered as an ideal
system for the Heisenberg model, but this is oversimplified. $4f$
wavefunctions in Gd have little overlap with $4f$ wavefunctions on
neighboring sites, so the direct exchange interaction between $4f$
states is almost zero. It is the $5d6s$ electrons that mediate the
intra-atomic exchange coupling among $4f$ electron spins.  $5d6s$
electrons are itinerant and across the Fermi level, rendering Gd
metallic and optically active.  In contrast to $3d$ transition metals,
where the same $3d4s$ electrons are responsible for both magnetic and
optical properties, in Gd $4f$ and $5d6s$ electrons are respectively
responsible for the magnetic properties and optical and transport
properties. The apparent disconnection between magnetic and optical
responses presents an opportunity to investigate intra-atomic exchange
coupling (IEC) in time domain.

Vaterlaus \et \cite{vaterlaus} carried out a time-resolved
spin-polarized photoemission and found that the spin-lattice
relaxation time in Gd is 100$\pm$80 ps. H\"ubner and Bennemann
\cite{wolfgang96} soon pointed out that the origin of this time scale
is the spin-orbit induced magnetocrystalline anisotropy energy and
showed a theoretical value of 48 ps, thus supporting the experimental
finding. However, Vaterlaus' pulse duration was too long to resolve
IEC. Beaurepaire and his coworkers \cite{eric} undertook an
unprecedented investigation of ultrafast spin dynamics in
ferromagnetic nickel films, thus opening a new frontier of
femtomagnetism \cite{ourreview,rasingreview,mplb16,walowski}.
Time-resolved second harmonic generation (SHG) was first employed to
detect the phonon-modulated coherent spin dynamics in Gd(0001)
ferromagnetic metal surfaces
\cite{melnikov2003,bovensiepen2004,melnikov2005}.  Lisowski \et
\cite{lisowski} performed the time-resolved photoemission measurement
and found that the spin polarization of the surface state is reduced
by half upon laser excitation, while the exchange splitting remains
unchanged. The linewidth of the surface states is broadened
\cite{loukakos}. A complete review on these earlier results is given
by Bovensiepen \cite{bovensiepen}. These SHG studies are useful for
surface states, but they do not have access to the $4f$ core level
(Fig. \ref{fig0}), thus IEC.  Melnikov \et \cite{melnikov2008b}
employed the magnetic linear dichroism to investigate the $4f$ core
level and showed that upon optical excitation of the $5d6s$ valence
electrons, the magnetic order in the $4f$ spin is reduced, from which
the intra-atomic exchange effect can be inferred.  Koopmans \et
\cite{koopmans2009} compared ultrafast demagnetization between Ni and
Gd through time-resolved magneto-optical Kerr effect.  Despite the
importance of IEC \cite{frietsch}, theoretical investigations have
been scarce, in sharp contrast to other studies
\cite{krieger,tows,freimuth}. Sandratskii \cite{sandratskii} did an
interesting calculation on the exchange splitting in surface and bulk
states by a static noncollinear configuration of $4f$ spins, but did
not investigate the IEC dynamics, neither did Oroszlany
\cite{oro}. Thus, a study of IEC is timely.

In this paper, we employ a time-dependent Liouville density functional
theory (TDLDFT) \cite{jpcm16} to study ultrafast inter-atomic exchange
in Gd.  We show that upon laser excitation, the $5d6s$ majority band
indeed shifts toward the Fermi level, while the minority band shifts
away from the Fermi level by 0.10 eV, in quantitative agreement with
the experiment \cite{carley,teichmann,frietsch}. The excited $5d6s$
electrons generate a new potential for otherwise optically silent $4f$
electrons, so $4f$ states are also shifted. In the many-body physics,
this corresponds to the intra-atomic exchange coupling, but is now
manifested in the time domain. We examine the occupancy at the
$\Gamma$ point before and after laser excitation and notice that
electrons even a few eV below the Fermi level are excited out of the
Fermi sea. However, at the energy window where $4f$ states appear,
there is no population loss. This proves that the effect on $4f$
states is indirect \cite{teichmann}. We scan along the $\Gamma$-M
direction or $\Sigma$ line in the reciprocal lattice space and notice
that the population loss is much stronger at the $M$ point than that
at the $\Gamma$ point.  We believe that our current study breaks new
ground by putting the theory at a semi-quantitative or quantitative
level, so a direct comparison between experiment and theory is now
possible.

The rest of the paper is arranged as follows. In Sect. II, we present
our theoretical formalism. The results and discussions are presented
in Section III. Finally, we conclude this paper in Section IV.

\section{Theoretical formalism}

\begin{figure}
  \includegraphics[angle=0,width=1\columnwidth]{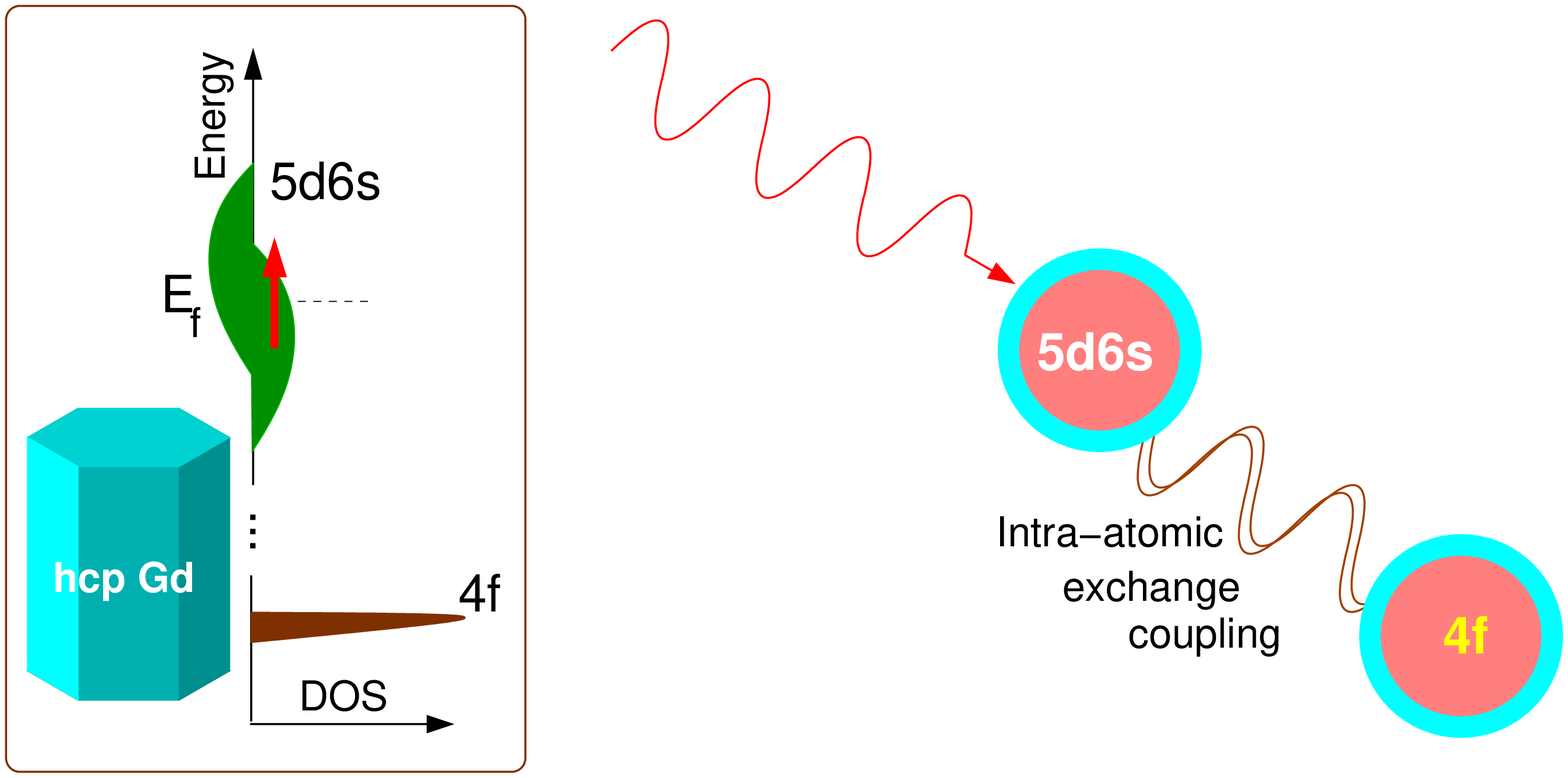}
  \caption{ (Left) Schematic of the electron density of states in hcp
    Gd.  $5d6s$ states are around the Fermi level and optically
    accessible, while $4f$ states are deeply below the Fermi energy
    and silent optically.  (Right) Laser pulses excites $5d6s$
    electrons first and, through the inter-atomic exchange coupling,
    affects highly localized $4f$ electrons.  }
\label{fig0}
  \end{figure}

Ground state properties of Gd have been thoroughly studied for a long
time. Gd has a standard hcp structure with lattice constants of
$a=3.636\rm \AA$ and $c=5.783\rm \AA$ (see Fig. \ref{fig0}) and with
two atoms per unit cell \cite{kurz}.  We directly use the experimental
lattice constants. The computed lattice parameters are very close to
the experimental ones \cite{kurz} and within 1\% (0.58\% for PBE
functional and 0.25\% with $+U$)\cite{petersen2006}.  Traditionally,
$4f$ states can be treated as core states or valence states, but Kurz
\et \cite{kurz} found that it is more accurate to treat it as valence
states. We employ the full-potential augmented plane wave method as
implemented in the Wien2k code \cite{wien}. We adopt the generalized
gradient approximation (GGA) for our density functional (Perdew, Burke
and Ernzerhof, 1996) \cite{pbe}, and include spin-orbit coupling
(SO). Both GGA+SO and GGA+SO+U calculations are performed, and our
results (density of states and magnetic moments) are fully consistent
with the prior investigations \cite{kurz,sabiryanov}. All the GGA+SO+U
results, $U$ and $J$ values, and other details are presented in the
supplementary materials.

To investigate laser-excited dynamics, we solve the Liouville equation
of motion for the density matrices $\rho$ at each $k$ point \cite{np09}
\begin{equation}
i\hbar \dot{\rho}=[H, \rho],
\end{equation}
where $H$ is the Hamiltonian and consists of two terms, one is for the
system and the other is for the interaction between the laser and
system. We choose the velocity gauge and the ${\bf p}\cdot {\bf A}(t)$
operator, where ${\bf p}$ is the momentum operator and ${\bf A}(t)$ is
the vector potential of the laser field in the unit of $\rm Vfs/\AA$.
We consider a circularly polarized laser pulse propagating along the
$-z$ axis (see Fig. \ref{fig0}). Our laser has a Gaussian shape with
duration $\tau=48$ fs and photon energy of $\hbar\omega=1.6$ eV. In
the traditional Liouville formalism \cite{prb09}, once the density is
obtained, one can directly compute the spin moment by tracing over the
product of the density matrix and spin operator. However, doing so
misses the important impact of the excited states on the system itself
and disregards the relaxation of the band structure \cite{jpcm15}. For
each time step, we feed the excited density back into the Kohn-Sham
equation and perform a self-consistent calculation under a constraint
excited potential where the electron occupation is held fixed. This
allows the excited state to create a new potential for the entire
system, so electrons, not directly excited optically, are
affected as well. This proves to be the key step to our method (for
details see Ref. \cite{jpcm15}). This also partially overcomes the
weak demagnetization in time-dependent density functional theory
\cite{krieger}. We should note that none of the current theories is
able to reproduce the same amount of the spin moment reduction under
the same experimental condition. Our TDLDFT theory represents a small
step forward by introducing a spin-scaling functional, so the excited
state spin information is fed back into the density \cite{mueller}.

\section{results and discussions}

\begin{figure}
  \includegraphics[angle=0,width=0.8\columnwidth]{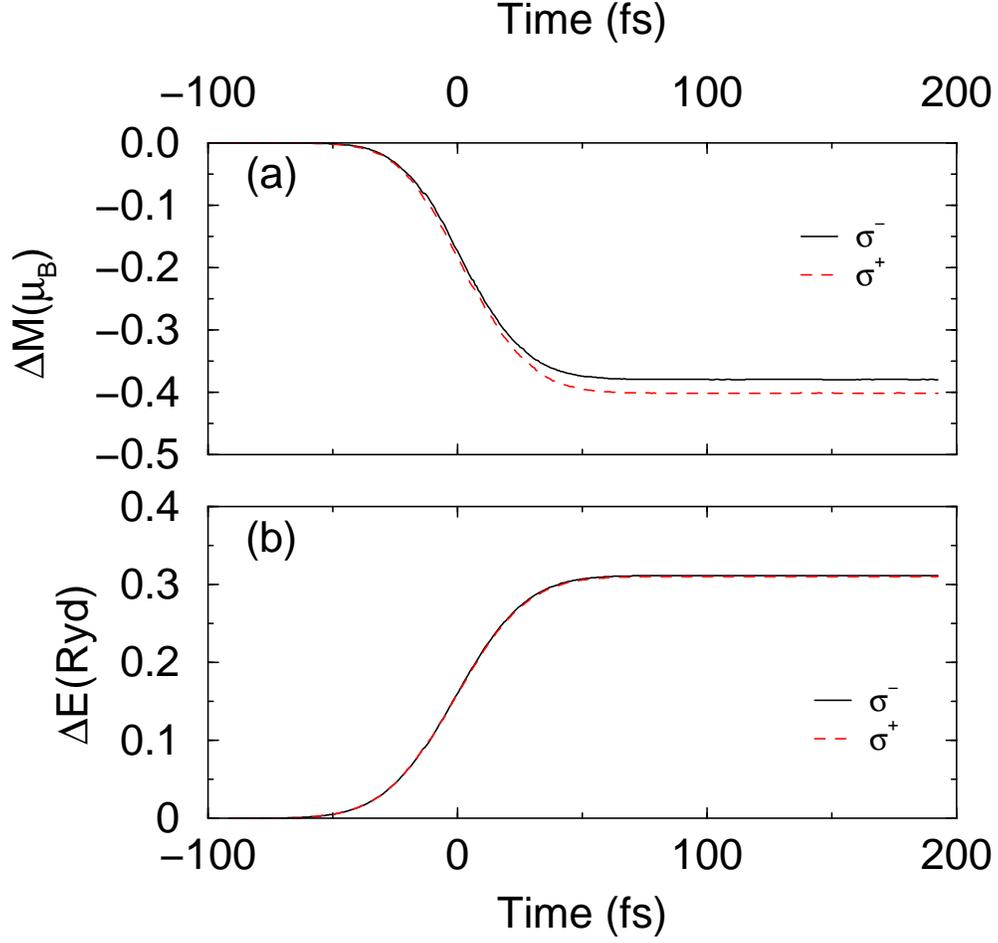}
  \caption{(a) Magnetic spin moment reduction $\Delta M$ as a function
    of time for left ($\sigma^-$, solid line) and right ($\sigma^+$,
    dashed line) circularly polarized light.  The laser pulse duration
    is 48 fs, its photon energy is 1.6 eV and the field amplitude is
    $0.03 \rm V fs/\AA$.  The spin minimum appears after the laser
    peaks. (b) Energy absorbed into the system. It follows the spin
    dynamics closely.}
\label{fig1}
  \end{figure}

\begin{figure}
  \includegraphics[angle=0,width=0.8\columnwidth]{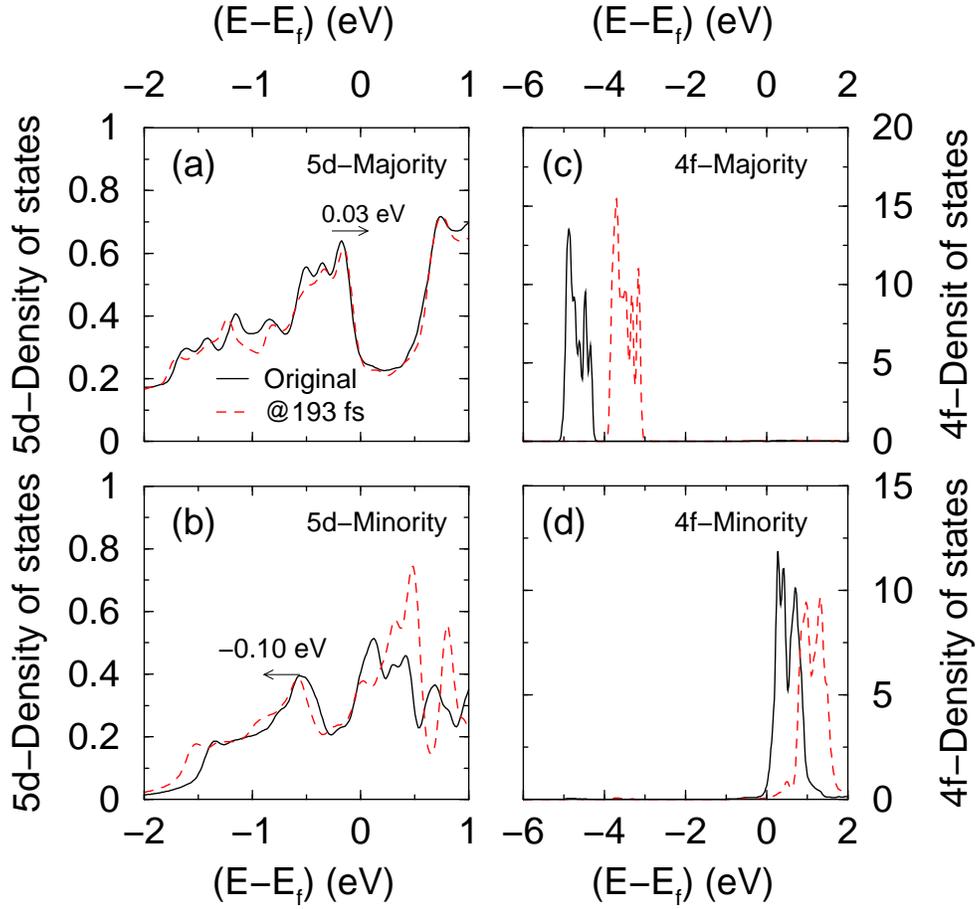}
  \caption{Densities of states (DOS) before (solid line) and after laser
    excitation (dashed line).  All DOS are computed at the GGA level
    and under $\sigma^+$ excitation. The laser field amplitude is
    $0.03 \rm Vfs/\AA$ and pulse duration is 48 fs.  The Fermi level
    is at 0 eV. (a) Upon laser
    excitation, $5d$-majority DOS is shifted toward the Fermi level by
    0.03 eV.  This is smaller than the experimental value
    \cite{carley}, but the trend is correct.  (b) $5d$-Minority spin
    DOS is shifted away from the Fermi level by -0.10 eV. This value
    quantitatively agrees with the experimental results \cite{carley}.
    (c) $4f$-Majority spin DOS is shifted toward the Fermi level by
    1.1 eV.  (d) $4f$-Minority spin DOS is shifted to the higher
    energy by 0.4 eV.  These two shifts need experimental verification.}
\label{d}
\label{f}
  \end{figure}

Figure \ref{fig1}(a) shows an ultrafast reduction in the spin moment
under left ($\sigma^-$) and right ($\sigma^+$) polarized light. The
laser helicity affects the amount of the reduction. $\sigma^+$ reduces
more due to the selection rule \cite{mplb16}. Figure \ref{fig1}(b)
shows the energy absorbed into the system during laser excitation. For
the same laser parameter, Gd absorbs more energy than fcc Ni. Since
our laser photon energy is 1.6 eV, the major excitation is within
$5d6s$ electrons around the Fermi level.  The fast response is mainly
due to the $5d$ electrons.  This can be seen clearly in the partial
density of states for $5d$ electrons in Fig. \ref{d}. Figure
\ref{d}(a) shows the majority spin states with a low binding energy
move toward the Fermi level by 0.03 eV (see the arrow), a trend that
is consistent with experimental findings
\cite{carley,teichmann,frietsch}. We set the Fermi level at 0 eV. Not
all the parts of the density of states behave similarly. Around -1 eV,
there is a clear modification in the structure of DOS. Quite
surprisingly, this is very similar to a latest report on the band
mirroring effect found in fct Co \cite{eich}, but this result comes
out of our first-principles calculation naturally, without invoking
other mechanisms.  Figure \ref{d}(b) shows that our minority band
moves away from the Fermi level by -0.10 eV, again consistent with the
experiments \cite{carley,teichmann,frietsch}. This shift is larger
than that for the majority state since the minority channel above the
Fermi level has a larger phase space and can easily receive electrons
while the majority channel has a big gap between 0 eV and 0.7 eV (see
Fig. \ref{d}(a)).

Our majority and minority band shifts should be compared with the
experimental results \cite{carley}. Carley \et \cite{carley} found
that the majority band shifts by 0.13 eV. This is much larger than our
theoretical results. Their minority band shifts by 0.097 eV and
matches our theoretical result almost quantitatively.  The difference
between our theory and their experiment is
understandable. Experimentally, the photoemission probes the exchange
along the $\Gamma$-M direction or $\Sigma$ line, but theoretically,
our results are from all directions; and the theoretical results
exactly along the $\Sigma$ line are difficult to obtain since our $k$
mesh is always slightly shifted.  It is likely that there is a
dispersion along different crystal momentum directions, a conjecture
that can be tested in future experiments.  To reduce the space
charging effect, the experimental pulse duration is stretched to 300
fs, but our theoretical duration is 48 fs, similar to Wietstruk \et
\cite{wietstruk} who used 50 fs.  In addition, Frietsch \et
\cite{frietsch2016} recently found that the exchange splitting depends
on the laser fluence. The stronger the fluence is, the larger the
shift in bands becomes. Considering these differences, the agreement
between the theory and experiment is very satisfactory and gives us
confidence in our method \cite{jpcm15}.

A central goal of our investigation is to understand how the
optically silent $4f$ states are changed through IEC during laser
excitation. Figures \ref{f}(c) and (d) compare the density of states
for $4f$ states before and after laser excitation.  A sharp narrow
peak, consisting of 7 electrons, is the hallmark of $4f$ states in Gd
\cite{kurz}.  Before laser excitation, the majority $4f$ states are
located around -4.5 eV below the Fermi level (see Fig. \ref{f}(c)),
while the minority band is empty and 0.5 eV above the Fermi
level. Because of this special energy arrangement, $4f$ electrons
cannot be directly excited optically in Gd, different from Tb
\cite{khorsand}. Figure \ref{f}(d) shows that at 193 fs after the
laser pulse peaks, both the majority and minority peaks are shifted to
a high energy side, and the partial density of states changes its
shape. The majority band shifts by 1.1 eV while the minority shifts
0.63 eV.  As a result the spin polarization is reduced. Wietstruk \et
\cite{wietstruk} detected this trend, but their data were noisy and
not accurate enough to make a comparison with our theory.  We should
point out that the large shift in $4f$ states is mainly due to the
overestimated itinerancy of the $4f$ states in the density functional
theory (both LDA and GGA levels) \cite{petersen2006}. The enhanced
$4f$ itinerancy increases the intra-atomic exchange interaction, so
the shift in the $4f$ state becomes larger.  The supplementary
materials show a smaller shift in $4f$ states under GGA+SO+U
approximation since $U$ term pushes the $4f$ states away from the
Fermi level and the intra-atomic interaction between $4f$ and $5d6s$
states is reduced.  Had we adopted the rigid band approximation, the
$4f$ energy level would never have been changed.

\begin{figure}
  \includegraphics[angle=0,width=0.8\columnwidth]{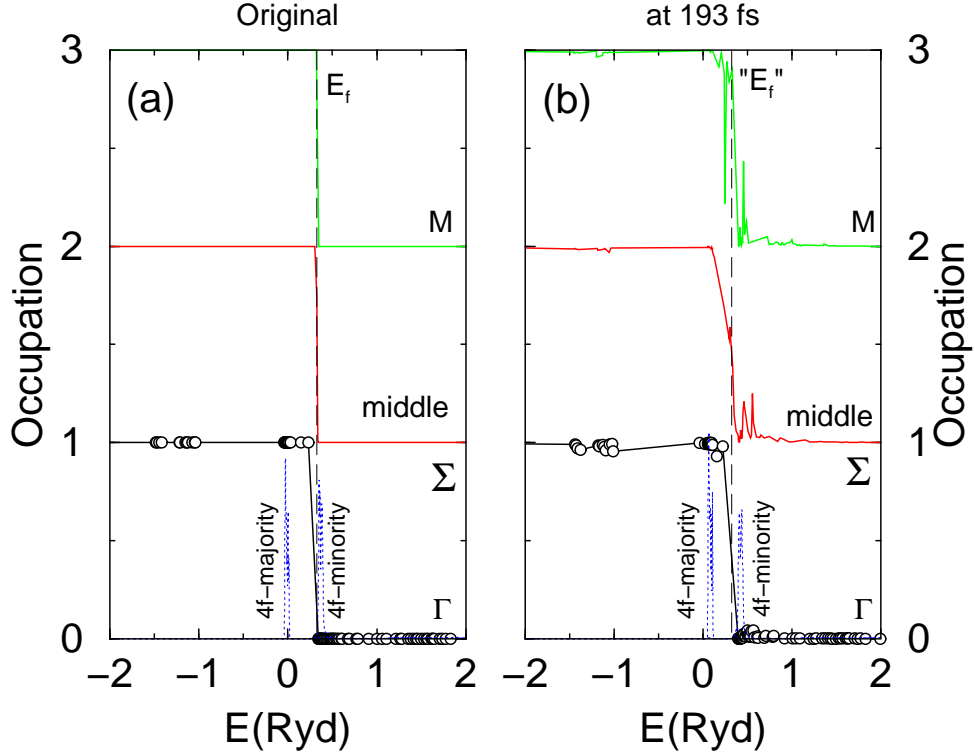}
  \caption{ Electron occupancy along the $\Gamma$-M direction
    ($\Sigma$ line). (a) Before laser excitation, the distributions
    are a typical Fermi-Dirac distribution.  Our $\Gamma$ point is
    approximate since our $k$ grid mesh is slightly shifted to improve
    convergence.  Different from Fig. \ref{f}, the energy scale is in
    Rydberg. The Fermi energy is not at 0; instead it is denoted by
    the vertical long-dashed line.  The occupancies for the middle
    point and M point are vertically shifted for clarity.  (Bottom
    inset) Both the majority and minority $4f$ partial densities of
    states are superimposed on the occupation (dotted line).  (b)
    After laser excitation, the electrons are excited out of the Fermi
    sea. The electron excitation is much stronger at the M point than
    at the $\Gamma$ point.  States several eV below the Fermi level
    are excited, but not at the $4f$ location.  }
\label{gamma}
\label{m}
  \end{figure}

We can prove that there is no direct excitation of $4f$
electrons. Figure \ref{gamma} illustrates the density of states along
the $\Sigma$ line (a) before and (b) after laser excitation. For
clarity, the occupations for the middle point and M point are
vertically shifted. We superimpose the $4f$-partial density of states
at the bottom of Figs. \ref{gamma}(a) and (b) so we can see where the
$4f$ electrons are located energetically. Note that this figure uses
Rydberg as its energy unit and the Fermi energy is not set at zero but
instead is denoted by a long-dashed line.  Before laser excitation,
all $k$ points have a normal Fermi distribution (Fig. \ref{gamma}(a))
and have no occupation above the Fermi level. At 193 fs after laser
excitation, the distribution function is non-Fermi-like, so our Fermi
energy is approximate. Figure \ref{gamma}(b) shows that electrons
several eV below the ``Fermi energy'' are excited out of the Fermi
sea, and excitation in each part of the occupation is non-uniform. The
electrons start to accumulate above the ``Fermi energy'' with a long
tail. In our calculation, we include 91 states covering 4 Rydberg, but
this may not be enough since we see that even original highest
unoccupied states have nonzero occupations. Importantly, at $4f$
states, there is no population loss (see Fig. \ref{gamma}(b)). The
entire distribution is no longer like a Fermi distribution. The change
becomes more pronounced as we move away from $\Gamma$ point. At the M
point, there is a big loss below the Fermi level, and electrons pile
up above the original Fermi level. We should add that there are
different methods to visualize the electron redistribution. An
interesting one is the crystal orbital overlap population method where
one plots the weighted charge distribution \cite{coop}. Another one is
to see how the hybridization changes upon laser excitation. 


Finally we address a key question: whether or not 
time-resolved photoemission
(TRPE) really detects a true magnetization change. From the
above comparison of the binding energy change, we see that our
theoretical value agrees with two experiments nearly
semi-quantitatively. This agreement could be fortuitous, but the fact
that both $\sigma^+$ and $\sigma^-$ light come to the same conclusion
suggests that our TDLDFT calculation catches important
physics. Additional evidence comes from the partial agreement between
our GGA+SO+U calculation and experiment (see the supplementary
material). Within our theory limit, we conclude that TRPE does probe
demagnetization.  This conclusion is conditional. First of all,
most TRPE probes emitted electrons only along one crystal momentum
direction.  If bands are narrow and flat and have little dispersion,
such as $d$ and $f$ bands, the spin change probed along one direction
in TRPE can be representative for the entire Brillouin zone.

Second, there is a crucial and unsettling difference in
demagnetization time between photoemission and magneto-optics and
magnetic dichroism.  In magneto-optics \cite{np09,sultan} and magnetic
dichroism \cite{wietstruk}, one probes the bulk magnetization
change. However, photoemission probes the spin polarization of the emitted
electrons (the number of spin up electrons minus the number of spin
down electrons), not the spin moment in the sample. In Gd
magneto-optics \cite{sultan} (MOKE) and magnetic circular dichroism
(MCD) \cite{wietstruk} agree on the demagnetization time of 750 fs
within the error margin of laser pulse duration. In TRPE, the time
constants of the binding energy shift are 200 fs and 900 fs for
minority and majority spins \cite{teichmann}. Neither of these 
matches 750 fs. However, the exchange splitting decreases within
860$\pm$100 fs \cite{carley}, which matches MOKE and MCD
demagnetization time. This suggests that the exchange splitting
reduction in TRPE, not the binding energy shift, is related to
demagnetization. This finding needs additional experimental
investigations.  Caution must be taken that one should not expect a
similar exchange splitting collapsing in rare-earth metals
\cite{andres2015} as that in transition metals since $4f$ electrons in
rare-earth metals strongly polarize $5d$ electrons.  This is
reminiscent of an  early study in Co \cite{aeschlimann1997}, where the spin-resolved
inelastic lifetime is as short as 20 fs, 10 times shorter than the
current established demagnetization time of 220 fs for fcc Co
\cite{radu2015}.


%
%
%
%
%
%


However, what determines demagnetization times in a sample has no
simple answer because both intrinsic and extrinsic factors play a
role. Current theories are unable to give a quantitative answer.  The
bulk demagnetization time of Gd is found around 0.7 ps
\cite{sultan2011} and increases by 10\% within a fluence change up to
1 mJ/cm$^2$.  Wietstruk \et \cite{wietstruk} showed that both Gd and
Tb have a similar ultrafast demagnetization time of 750 fs with an
uncertainty of 250 fs.  Koopmans \et \cite{koopmans2009} suggested a
simple expression that relates the demagnetization time to the Curie
temperature $T_c$ and magnetic moment $\mu_{at}$, $T_c/\mu_{at}$ of a
sample. However, when Wietstruk \et \cite{wietstruk} applied it to Gd
and Tb, they could not explain a similar demagnetization time in Gd
and Tb \cite{wietstruk}.  Our above simulation singles out the
importance of excited charge density in demagnetization.  Both Gd and
Tb involve the same type of $5d6s$ electrons.  We propose an
alternative relation that the demagnetization time should be
proportional to \be \tau_m \propto \frac{1}{\rho(E_f)\mu_{x}},\ee
where $\mu_x$ refers to the spin moment of the {\sl excited
  electrons}, not all the electrons, and $\rho(E_f)$ is the density of
states at the Fermi level.  Quantitatively, the $3d$ density of states
in Ni at its Fermi level is about 1.5 states/eV \cite{jpcm16} and the
spin moment is 0.6$\mu_B$, while in Gd, the $5d$ density of states in
Gd is about 0.3 states/eV and the spin moment of $5d$ is about 0.58
$\mu_B$. According to Mathias \et \cite{mathias}, the demagnetization
time for Ni is 157 fs, so this gives the demagnetization time for Gd,
785 fs. This differs from Wietstruk's results by 35 fs, well within
their pulse duration of 100 fs \cite{stamm}.  Given that these two
experiments use entirely different techniques, this quantitative
agreement is truly gratifying, and should be tested in other
systems. Physically, the above expression appears to be more
reasonable since it is consistent with the basic theory of fundamental
excitation in metals.  If there is low density of states at the Fermi
level, the relaxation among the electrons themselves is going to be
slow. In half-metals such as CrO$_2$, one channel is shut off, so the
relaxation must be slow \cite{zhang2002}.

Very recently, Frietsch \et \cite{frietsch} reported two different
dynamics for $5d$ (800 fs) and $4f$ electrons (14 ps). This is fully
expected since $4f$ electrons have nearly zero density of states at
the Fermi level and are indirectly excited through the intra-atomic
exchange interaction. The short time dynamics is associated with the
$5d6s$ electrons.  While an extensive study on this is beyond the
scope of this paper, we propose to measure the density of states in
those prior samples and reexamine those experimental results
\cite{mann}.

\section{conclusion}

Our time-dependent first-principles calculation has shed new light on
how intra-atomic exchange correlation develops in Gd after laser
excitation.  It starts with the $5d6s$ electrons because of their
proximity to the Fermi level.  Because of differences in their phase
spaces, the majority and minority $5d6s$ electrons respond
differently: The majority band shifts toward the Fermi level, while
the minority moves away from the Fermi level. Our theory now agrees
with the experimental results \cite{carley} qualitatively and even
quantitatively for the minority bands. To the best of our knowledge,
this has never been attempted.  The excited $5d6s$ electrons not only
affect themselves, but also generate a new potential for optically
inaccessible $4f$ electrons, so $4f$ states feel the impact of laser
excitation. We find that the $4f$-density of states changes its
original shape, also seen in Co \cite{eich}, and shifts its position
as well, a prediction that must be tested experimentally.  Although
$5d6s$ electrons are excited intensively, there is no change at $4f$
states, a hallmark of the intra-atomic exchange correlation.  Our
study represents a beginning and is expected to have a broad impact on
future research. In particular, nearly a quantitative agreement
between the density functional calculation and the experiments allows
an unbiased comparison. This will certainly encourage new theoretical
and experimental efforts in ultrafast demagnetization in rare-earth
materials.

\acknowledgments We would like to thank Prof. P. Blaha of Technical
University of Wien for helpful communications on the GGA+SO+U
calculation, and Prof. Uwe Bovensiepen and Dr. Andrea Eschenlohr of
University Duisburg-Essen for fruitful discussions and for the
information on Ref. \cite{radu2015}.  This work was solely supported
by the U.S. Department of Energy under Contract
No. DE-FG02-06ER46304. Part of the work was done on Indiana State
University's quantum cluster and high-performance computers.  The
research used resources of the National Energy Research Scientific
Computing Center, which is supported by the Office of Science of the
U.S. Department of Energy under Contract No. DE-AC02-05CH11231.

$^*$gpzhang.physics@gmail.com

\end{document}